\documentclass[aps,11pt,showpacs,showkeys]{revtex4}
\usepackage[dvips]{graphicx}

\begin{document}
\title{Are there crystal field levels in UPd$_{2}$Al$_{3}?$ We answer, THERE ARE$^\spadesuit$}
\author{R. J. Radwanski}
\homepage{http://www.css-physics.edu.pl}
\email{sfradwan@cyf-kr.edu.pl}
\affiliation{Center of Solid State Physics, S$^{nt}$Filip 5, 31-150 Krakow, Poland,\\
Institute of Physics, Pedagogical University, 30-084 Krakow, Poland}
\author{Z. Ropka}
\affiliation{Center of Solid State Physics, S$^{nt}$Filip 5,
31-150 Krakow, Poland}

\begin{abstract}
We claim that crystal field (CEF) levels exist in UPd$_2$Al$_3$ in
contrary to a recent claim of Hiess {\it et al.}
(cond-mat/0411041) and Bernhoeft {\it et al.} (cond-mat/0411042),
that there is no experimental evidence for discrete crystal field
levels in this superconducting heavy-fermion antiferromagnet. We
claim that excitations revealed by Krimmel {\it et al.} (J. Phys.:
Condens. Matter \textbf{8} (1996) 1677) in
inelastic-neutron-scattering (INS) studies are i) crystal-field
excitations described by us within ii) the 5f$^{3}$ (U$^{3+}$)
configuration. Moreover, our 5f$^{3} $(U$^{3+}$) scheme, presented
in Physica B \textbf{276-278} (2000) 803 and in Czech. J. Phys.
\textbf{54} (2004) D295, provides a clear physical explanation for
the 1.7 meV excitation (magnetic exciton) as associated to the
removal of the Kramers-doublet ground state degeneracy in the
antiferromagnetic state. The crystal-field theory completed by
strong intra-atomic correlations and intersite spin-dependent
interactions to the Quantum Atomistic Solid State Theory (QUASST),
offers the meV energy scale needed for description of magnetic and
electronic properties of compounds containing open-shell 3d, 4f,
5f atoms. The derived set of CEF parameters for the U$^{3+}$ state
reproduces both the INS excitations, temperature dependence of the
heat capacity, large uranium magnetic moment as well as its
direction.

\end{abstract}
\maketitle\vspace {-1.0cm}

\section{Introduction}
\vspace {-0.5cm}The description of electronic and magnetic
properties of UPd$_2$Al$_3$ is still under hot debate though its
exotic properties has been discovered already more than 10 years
ago \cite{1,2}. The uniqueness of UPd$_2$Al$_3$ relies in the
coexistence of the heavy-fermion (h-f) phenomena, the large
magnetic moment of about 0.85-1.5 $\mu _B$ in the antiferomagnetic
state below T$_N$ =14.3 K and the superconductivity below 2 K. The
main point of this debate is related to the understanding of the
role played by f electrons.

In band calculations f electrons are considered as itinerant
\cite{3,4,5,6,7} whereas surprisingly nice reproduction of many
experimental results can be obtained within the crystal-field
(CEF) approach \cite{2,8,9,10,11,12}, i.e. treating f electrons as
localized. Within the CEF approach there is presently discussion
about the tetravalent \cite{2,8,10,11} or trivalent \cite{9,12}
uranium state in UPd$_2$Al$_3$. When it have seemed that the
problem of the existence of the crystal field levels in
UPd$_{2}$Al$_{3}$ has come to the positive conclusion (Fulde and
Zwicknagl by years propagating the itinerant behaviour of f
electrons as the origin of heavy-fermion phenomena have admitted
in 2002 \cite{13,14}) the existence in UPt$_{3}$ and
UPd$_{2}$Al$_{3}$ of two localized f electrons with the f$^{2}$
multiplet structure and after seemed to unambiguous experimental
evidence of crystal field excitations due to Krimmel {\it et al.}
\cite{15} recently appear two papers by Hiess {\it et al.}
cond-mat/0411041 \cite{16} and by Bernhoeft {\it et al.}
cond-mat/0411042 \cite{17} with a clear and sharp negative
statement about crystal field levels in UPd$_{2}$Al$_{3}$. The
appearance of these two papers is the direct motivation for this
paper. Here we would like to clarify a problem, "Are there
crystal-field levels in UPd$_{2}$Al$_{3}$?". Our answer is, THERE
ARE. However, as we do not think that so prominent physicists like
Prof. Prof. N. Aso, N. Bernhoeft, Y. Haga, A. Hiess, Y. Koike, T.
Komatsubara, G.H. Lander, N. Metoki, Y. Onuki, B. Roessli, and
N.K. Sato do not know what crystal-field levels are by this paper
we would like to start an open discussion about the crystal field
theory and its physical adequacy. The biggest problem in this
discussion is related to a fact that the crystal-field theory has
within the magnetic community in last 30 years a special place -
being continuously rejected from the scientific life permanently
appears as a unavoided approach for explanation of properties of
real compounds. The crystal-field theory is in the modern
solid-state theory like a unwilling child, 75 years old already.
\vspace {-0.7cm}
\section{$\textrm{f}$  states  in  UP$\textrm{d}_{2}$A$\textrm{l}_{3}$ - a historical outline}
\vspace {-0.6cm}We are, in particular R.J. Radwanski, by years
formulating the fundamental controversy in the theoretical
understanding of compounds containing open-electron shells (3d, 4f
and 5f), including those exhibiting heavy-fermion phenomena, as
related to the treatment of f electrons (as localized or
itinerant) and subsequently the formation by them a wide 2-5 eV
band or low-energy discrete crystal-field states, in the meV
scale. In particular, in 1992, when the itinerant f electron
picture, with the f electrons put at the Fermi level, was fully
dominating I (RJR) came out with a view that the heavy-fermion
phenomena are related to crystal-field interactions of a
strongly-correlated odd-number electron system \cite{18,19} like
the Ce$^{3+}$ (5f$^{1}$), Yb$^{3+}$ (4f$^{13}$) and U$^{3+}$
(5f$^{3}$) configuration (all are Kramers ions; it is worth
noting, that in fact, the U$^{3+}$ ion itself is a heavy fermion
system with strongly-correlated atomic-like 89 electrons) and, for
UPd$_{2}$Al$_{3}$ we have explained the overall temperature
dependence of the heat capacity, including the $\lambda$-type peak
at T$_{N}$, as the crystal-field contribution of the U$^{3+}$
(5f$^{3}$) configuration \cite{9}. At the same Conference Frank
Steglich's group have presented CEF calculations for the U$^{4+}$
(5f$^{2}$) configuration \cite{8}. The CEF approach, completed
with inter-site spin-dependent interactions, we have successfully
applied for description of electronic and magnetic properties of
conventional rare-earth intermetallics (ErNi$_{5}$, NdNi$_{5}$,
Ho$_{2}$Co$_{17}$, Nd$_{2}$Fe$_{14}$B, ...) \cite{20,21,22},
though up to now there is strong opposition for the use of the
crystal-field approach \cite{23} to intermetallic compounds. At
that time we have extended the crystal-field theory to an
individualized electron model for rare-earth intermetallics, which
subsequently has been extended to Quantum Atomistic Solid State
Theory (QUASST) for transition-metal 3d-/4f-/5f-atom containing
compounds \cite{24}. In 1996 Krimmel {\it et al.} have presented
their results of "Search for crystal-field excitations in
UPd$_{2}$Al$_{3}$" \cite{15} undertaken with an aim to confirm
their 5f$^{2}$ configuration. They did not manage to attribute the
observed excitations to the 5f$^{2}$ scheme similarly as a later
paper by Schenck {\it et al.} \cite{11}. In a year of 2000, in our
paper, Ref. 12, we took experimental results of Krimmel {\it et
al.} as confirmation of the existence of crystal-field states in
UPd$_{2}$Al$_{3}$, exactly in the same line as Krimmel {\it et
al.} have presented own results (despite a fact, that there is the
controversy about a detailed description of the observed states:
are they related to 5f$^{2}$ or 5f$^{3}$ configuration, what we
treat as an important but a minor problem within the
localized-electron paradigm) and we have attributed the observed
excitations to the strongly-correlated electron f$^{3}$ scheme
proposed by us already in 1992. According to us, and we thought up
to now that for everybody, the INS experiment of Krimmel et al.
\cite{15} has revealed at 25 K crystal-field excitations with
energies of 7 and 23.4 meV. This experiment at 150 K has revealed
further excitations at 3 and 14 meV at the energy-loss side and at
7 meV at the energy-gain side.

Sato, Steglich {\it et al.} in 2001 \cite{25,26} and later Fulde
and Zwicknagl \cite{13,14} accepting the existence of the
localized levels of Krimmel {\it et al.} came out with a concept
of a fragmentation of the 5f shell into a local moment and
itinerant state known at present as the dual nature of f
electrons. The local 5f$^{2}$ subshell should manifest in the
multiplet structure with the $\Gamma _{4}$ ground state
originating from:

$\Gamma _{3,4}$ = $\frac{1}{\sqrt{2}}$ ($|$J=4;J$_{z}$=+3$>$ $\pm$
$|$J=4;J$_{z}$=-3$>$)

split in energy, of $\sim$7 meV, by the crystalline electric field
\cite{14}. Although these authors did not present a detailed
analysis for the confirmation of the 5f$^{2}$ multiplet structure
they obviously think of in terms of the crystal field theory. The
recalled energy splitting is the excitation observed by Krimmel
{\it et al.} \cite{15}. In Ref. \cite{25} an extra low-energy
excitation of 1.7 meV, at T = 0 K, has been established to exist
in the magnetically ordered state and this excitation has been
ascribed by authors of Ref. \cite{25} to a magnetic exciton, with
the value determined by uniaxial exchange anisotropies. In the
superconducting state an ultra-low energy excitation of 0.35 meV
at T = 0 K appears.

In such experimental and theoretical situation we found with a
really big surprise a recent claim of Hiess {\it et al.} and
Bernhoeft {\it et al.} that in UPd$_{2}$Al$_{3}$ there is no
evidence for CEF levels. Below we cite in the full extension their
statements:

Hiess {\it et al.} \cite{16} - in Resume on p. 4 write: "Work on
polycrystalline material at the ISIS spallation source by Krimmel
{\it et al.} then followed giving an overview of the inelastic
response function up to ~20 meV. This study gives no evidence for
a discrete crystal field level scheme and ...."

Bernhoeft {\it et al.} \cite{17} on p. 12: "Furthermore, the
fundamental assumption of strongly localised 5f levels is
difficult to reconcile with the lack of observation of CEF levels
in the paramagnetic state [here citation to Krimmel {\it et al.},
Ref. \cite{15}] and the success of ab-initio band-structure
calculations using the delocalised LSDA approach to reproduce
experimental Fermi surface areas, as measured by dHvA effect,
which often are taken as an indication that the 5f levels are
largely delocalised."(here is citation to Knopfle {\it et al.},
Ref. \cite{5} and to Inada {\it et al.}, Ref. \cite{7}.)

Thus, according to authors of the cond-mat/0411041 paper: A.
Hiess, N. Bernhoeft, N. Metoki, G.H. Lander, B. Roessli, N.K.
Sato, N. Aso, Y. Haga, Y. Koike, T. Komatsubara, and Y. Onuki, and
the cond-mat/0411042 paper : N. Bernhoeft, A. Hiess, N. Metoki,
G.H. Lander and B. Roessli, there is no evidence for discrete
crystal field levels in UPd$_{2}$Al$_{3}$. We claim that such a
sharp negative statement is completely wrong. It is also, as we
have shown above, in the sharp contrast with the interpretation of
Krimmel {\it et al.}'s results of previously recalled researchers.
We do not agree that there is "the lack of observation of CEF
levels in the paramagnetic state" as the INS experiments conducted
at 25 K and 150 K are both in the paramagnetic state.

Such opinion of prominent physicists introduces an enormous chaos
in the general understanding of the magnetism and the electronic
structure of open-shell compounds, here of UPd$_{2}$Al$_{3}$. We
claim that properties of UPd$_{2}$Al$_{3}$ can be only understood
starting from localized states of f electrons, in a number of 2 or
3. We prefer 3 f electrons owing to the intrinsic dynamics of the
Kramers system, states of which are established by the atomic
physics (in particular the number of states and their
many-electron atomic-like nature). The detailed electronic
structure is predominantly determined by conventional interactions
in a solid: the Stark-like effect by the crystalline electric
field potential due to 3-dimensional array of charges in a crystal
acting on the aspherical incomplete shell, and the Zeeman-like
effect due to spin-dependent interactions of the incomplete-shell
spin (atomic-like moment) with self-consistently induced spin
surroundings. These states can become broaden in energy by
different interactions (lowering of the local symmetry, thermal
expansion, appearance of a few inequivalent sites, lattice
imperfection, surface effects and other solid-state effects).
Obviously, we should not think that discrete crystal field states
mean that they are extremely thin lines. 3 or even 10 meV broad
lines are still of the crystal-field origin. Underlying by us by
many, many years the importance of the crystal field we have
treated as an opposite view to the overwhelmed band structure view
yielding the spreading of the f-electron spectrum by 2-5 eV.

\begin{figure}[ht]
\begin{center}
\includegraphics[width = 9.5 cm]{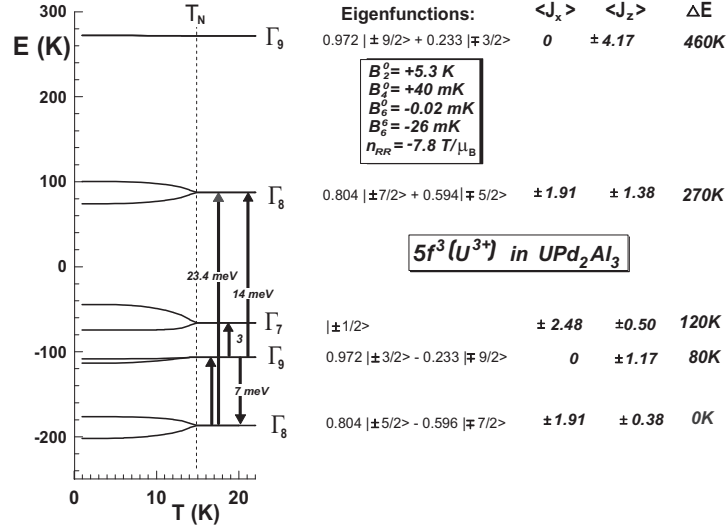}
\end{center}
\vspace {-0.8cm}\caption{Energy level scheme of the U$^{3+}$ ion
in UPd$_2$Al$_3$ taken from our papers \cite{12,27}. Arrows
indicate transitions which we have attributed to excitations
revealed by inelastic-neutron-scattering experiments of Krimmel
{\it et al.} \cite{15}.}
\end{figure}

In Ref. \cite{12} (and recently in Ref. \cite{27}; a direct
motivation for this paper was a claim \cite{13,14,25} for the
5f$^{2}$ configuration) we have interpreted the excitations of
Krimmel {\it et al.} as related to the energy level scheme: 0, 7
meV (81 K), 10 meV (116 K) and 23.4 meV (271 K) and we have
ascribed this scheme to the $f^{3}$ (U$^{3+}$) scheme. The fine
electronic structure of the $f^{3}$ (U$^{3+}$ consists of five
Kramers doublets split by multipolar charge interactions (CEF
interactions). The fine electronic structure originates from the
lowest multiplet $^{4}$I$_{9/2}$, higher multiplets are at least
0.2 eV above and do not affect practically the ground-multiplet
properties. These 5 doublets are further split in the
antiferromagnetic state, i.e. below T$_N$ of 14 K in case of
UPd$_2$Al$_3$ as is shown in Fig. 1. A derived set of CEF
parameters of the hexagonal symmetry: B$_2^0$=+5.3 K, B$_4^0$=+40
mK, B$_6^0$=-0.02 mK and B$_6^6$=-26 mK yields states at 81 K, 120
K, 270 K and 460 K, energies of which are in perfect agreement
with the experimentally observed excitations. The highest state
was not observed in INS experiment. This electronic structure
accounts also surprisingly well for the overall temperature
dependence of the heat capacity, the substantial uranium magnetic
moment and its direction if we make use of a single-ion like
Hamiltonian for the ground multiplet J=9/2 \cite{12}:

\begin{center}
$H=H_{CF}+H_{f-f}=\sum \sum B_n^mO_n^m+n_{RR}g^2\mu _B^2\left(
-J\left\langle J\right\rangle +\frac 12\left\langle J\right\rangle
^2\right) $
\end{center}

The first term is the crystal-field Hamiltonian. The second term
takes into account intersite spin-dependent interactions (n$_{RR}$
- molecular field coefficient, $g$=11/8 - Lande factor) that
produce the magnetic order below T$_N$ what is seen in Fig. 1 as
the appearance of the splitting of the Kramers doublets and in
experiment as the $\lambda$-peak in the heat capacity at T$_{N}$.
In Fig. 2 we have shown temperature dependence of the splitting
energy between two conjugate Kramers ground state. Surprisingly,
both the value of the energy and its temperature dependence is in
close agreement to a low-energy excitation of 1.7 meV at T=0 K
observed by Sato {\it et al.} \cite{25} which has been attributed
by them to a magnetic exciton. The agreement is very remarkable
taking into account that Fig. 2 has been made by the simple
subtraction of two lowest states shown in our published papers
\cite{27,12}. Thus, we are convinced that the 5f$^{3} $(U$^{3+}$)
scheme provides a clear physical explanation for the 1.7 meV
excitation (magnetic exciton) - this excitation is associated to
the removal of the Kramers-doublet ground state degeneracy in the
antiferromagnetic state.
\begin{figure}[ht]
\begin{center}
\includegraphics[width = 7.2 cm]{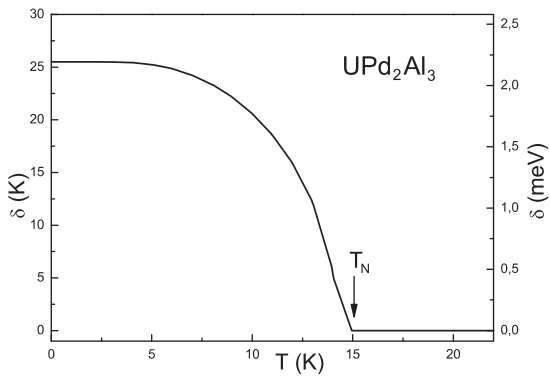}
\end{center}
\vspace {-0.8cm}\caption{Temperature dependence of the excitation
energy $\delta$ to the Kramers conjugate state of the $f^{3}$
(U$^{3+}$) configuration in UPd$_2$Al$_3$. This energy is obtained
as the difference between energies of two lowest states seen in
Fig. 1, taken from Ref. \cite{27}. Both the value of the energy
and its temperature dependence is in close agreement to a
low-energy excitation of 1.7 meV at T = 0 K observed by Sato {\it
et al.} \cite{25} and attributed by them to a magnetic exciton.}
\end{figure}
\vspace {-0.8cm}
\section{Crystal-field theory - some remarks}
\vspace {-0.2cm} A strange scientific climate about the
crystal-field theory in the modern solid-state paradigm comes from
a widely-spread view within the magnetic community that it does
not have the proper theoretical justification. An oversimplified
point charge model is treated as an essence of the crystal-field
theory. An indication in some cases that the point charge model is
not sufficient to account for the crystal-field splittings was
taken as the conclusive proof for the incorrectness of the
crystal-field concept at all. According to us the theoretical
background for the crystal-field theory is the atomic construction
of matter. Simply, atoms constituting a solid preserve much of
their atomic properties. One can say that the atomic-like
integrity is preserved, after giving up partly or fully some
electrons, and then the atomic identity serves as the good quantum
number of the electron system. We are quite satisfied that the
point-charge model provides the proper variation the ground states
going on from one to another 3d/4f/5f atom. Different ionic states
we consider as different states of the atom, though it is better
instead of the ionic state to say about the electron
configurations and their different contributions to magnetic,
electronic, spectroscopic and optical properties. For instance, in
metallic ErNi$_{5}$ there exists 4f$^{11}$ electron configuration,
often written as the Er$^{3+}$ ion, that is found to be
predominantly responsible for the magnetism and the electronic
structure of the whole compound \cite{20,21,22}; the other
electrons of Er and Ni are responsible for the metallic behavior.
We point out the multipolar character of the electric potential in
a solid. It is very fortunate situation when a solid, with
milliard of milliard of atoms, can be described with the single
electronic structure. It is true, that the crystal-field theory
being itself a single-ion theory cannot describe a solid with
collective interactions. For this reason we came out with the
Quantum Atomistic Solid State Theory and completed the
crystal-field theory with strong intra-atomic correlations and
intersite spin-dependent interactions. By pointing out the
importance of the CEF theory we would like to put attention to the
fundamental importance of the atomic physics (Hund's rules,
spin-orbit coupling, ....) and local single-ion effects. It is
worth remind that the source of a collective phenomenon, the
magnetism of a solid, are atoms constituting this solid.
Properties of these potentially-active atoms (open-shell atoms)
are determined by local surroundings and local symmetry.
Subsequently, these atomic moments, with spin and orbital parts,
enter to the collective game in a solid.

Another wrong conviction about the crystal-field theory is that it
was exploited already completely. In order to shown that this
thinking is wrong we turn the reader's attention that the
crystal-field approach used within the rare-earth and actinide
community (4f and 5f systems) fundamentally differs from that used
within the 3d community. The 4f/5f community works with $J$ as the
good quantum number whereas the 3d community "quenches" the
orbital moment and works with only the spin $S$. Our description
of a 3d-atom compound like FeBr$_{2}$ and LaCoO$_{3}$ one can find
in Refs \cite{28} and \cite{29}. In case of the
strongly-correlated crystal-field approach we work with
many-electron states of the whole 4f$^{n}$, 5f$^{n}$ 3d$^{n}$
configuration in contrary to single-electron states used in 3d
magnetism and LDA, LSDA, and many other so-called {\it ab initio}
approaches. Technically, strong correlations are put within the
CEF theory, and in QUASST, by application of two Hund's rules. The
{\it ab initio} calculations will meet with the CEF (QUASST)
theory in the evaluation of the detailed charge distribution
within the unit cell and after taking into account strong
intra-atomic correlations among electrons of incomplete shells and
the spin-orbit coupling in order to reproduce the CEF conditions
(two Hund's rules, also the third one for rare-earths and
actinides).

We would like to mention that we are fully aware that used by us
the Russell-Saunders LS coupling can show some shortages in case
of actinides related to the growing importance of the j-j
coupling. We are aware of many other physical problems which we
could not mention here due to the length problem - finally we
mention only that we can reverse scientific problem in the solid
state physics and use 4f/5f/3d compounds as a laboratory for the
atomic physics for study 3d/4f/5f atoms in extremal electric and
magnetic fields. In the solid-state physics we study the lowest
part of the atomic structure but extremely exactly.

We call for an open and honest scientific discussion on the
strongly-correlated crystal-field theory being convinced that it
is the fundamental ingredient of the modern solid-state paradigm.
There is a hope that such discussion can proceed in Phys. Rev. B
where the Hiess {\it et al.} (cond-mat/0411041) and Bernhoeft {\it
et al.} (cond-mat/0411042) have submitted their papers and after
being printed we intend to comment them.

\section{Conclusions}
\vspace {-0.5cm} We claim that excitations revealed by Krimmel
{\it et al.} (J. Phys.: Condens. Matter \textbf{8} (1996) 1677) in
inelastic-neutron-scattering (INS) studies of superconducting
heavy-fermion antiferromagnet UPd$_2$Al$_3$ are i) crystal-field
(CEF) excitations described by us within ii) the 5f$^{3}$
(U$^{3+}$) configuration, in contrary to a recent claim of Hiess
{\it et al.} (cond-mat/0411041) and Bernhoeft {\it et al.}
(cond-mat/0411042), that there is no experimental evidence for
discrete crystal-field levels. Moreover, our U$^{3+}$ scheme,
presented in Refs \cite{12} and \cite{27}, provides a clear
physical explanation for the 1.7 meV excitation (a magnetic
exciton) as associated to the removal of the Kramers-doublet
ground state degeneracy in the antiferromagnetic state. The
crystal-field view completed by strong intra-atomic correlations
and intersite spin-dependent interactions to the Quantum Atomistic
Solid State Theory (QUASST), offers the meV energy scale needed
for description of magnetic and electronic properties of compounds
containing open-shell 3d, 4f, 5f atoms. The derived set of CEF
parameters for the U$^{3+}$ state in UPd$_2$Al$_3$ reproduces both
spectroscopic and thermodynamical properties and a substantial
uranium magnetic moment as well as its direction. In metallic
UPd$_2$Al$_3$ three localized f electrons coexist with conduction
electrons originating from outer 7s$^{2}$/6d electrons of uranium
as well as from Pd and Al outer electrons \cite{27}. Our studies
point out the necessity of discussion of electronic and magnetic
properties on the atomic scale, where the local symmetry, crystal
field and strong electron correlations, of the intra-atomic and
inter-site origin, play the fundamental role in the formation of
the low-energy discrete energy spectrum. Our localized view is
supported by results of Fujimori {\it et al.} \cite{30} who found
that the resonant photoemission spectrum of UPd$_{2}$Al$_{3}$
reflects the single site effect of uranium sites. We are convinced
that strongly-correlated crystal-field theory is a fundamental
ingredient of the modern solid-state paradigm.

We do not claim to (re-)invent crystal-field theory but we
advocate for its high physical adequacy as a basic ingredient of
the modern solid-state paradigm. We call for an open and honest
scientific discussion on the strongly-correlated crystal-field
theory as the fundamental ingredient of the modern solid-state
paradigm. We understand our scientific work as continuation of
works of J. H. Van Vleck, H. A. Kramers, K. W. H. Stevens and
many, many others started in 1929 by Hans Bethe. \\
$^\spadesuit$Dedicated to H. Bethe, J. H. Van Vleck and K. W. H.
Stevens on the 75$^{th}$ anniversary of the crystal-field theory.

\end{document}